\documentclass[english,12pt]{article}
          \usepackage{babel}
          \usepackage[T1]{fontenc}
          \usepackage[latin2]{inputenc}
          \usepackage{pslatex}
\begin{document}
\begin{flushright}
\textit{ACTA ASTRONOMICA}\\
Vol.0 (2014) pp.0-00
\end{flushright}
\bigskip

\begin{center}
{\bf CzeV404 - an eclipsing dwarf nova in the period gap\\
during its July 2014 superoutburst}\\
\vspace{0.8cm}
{K. ~B~\k{a}~k~o~w~s~k~a$^{1}$, A. ~O~l~e~c~h$^1$, R. ~P~o~s~p~i~e~s~z~y~\'{n}~s~k~i$^2$, \\
F. ~M~a~r~t~i~n~e~l~l~i$^3$ \& A. ~M~a~r~c~i~n~i~a~k$^4$}\\
\vspace{0.5cm}
\begin{small}
{$^1$ Nicolaus Copernicus Astronomical Center,
Polish Academy of Sciences,\\
ul.~Bartycka~18, 00-716~Warszawa, Poland\\
$^2$ Comets and Meteor Workshop,
ul.~Bartycka~18, 00-716~Warszawa, Poland\\
$^3$ Lajatico Astronomical Centre, Loc i Fornelli N$^{o}$9, Orciatico Lajatico, Pisa, Italy\\
$^4$ Astronomical Observatory Institute, Faculty of Physics, 
A.Mickiewicz University,\\ ul. S{\l}oneczna 36, 60-286 Pozna\'{n}, Poland\\}
\end{small}
{\tt e-mail: bakowska@camk.edu.pl}\\
~\\
\end{center}

\begin{abstract}  

Results of the CCD observations of CzeV404 are displayed. During the
season of June-August 2014 we detected one outburst and one
superoutburst of the star. Clear superhumps with the period of $P_{sh}=0.10472(2)$
days were observed. The superhump period was decreasing with a high
value of \textit{\.{P}}=$-2.43(8) \times 10^{-4}$. From 17 eclipses we
calculated the orbital period with the value of $P_{orb}=0.0980203(6)$ days
which confirms that CzeV404 belongs to period gap objects and it
is the longest orbital period eclipsing SU UMa star. Based on
superhump and orbital period determinations, the period excess
$\varepsilon=6.8\% \pm 0.02 \%$ and the mass ratio $q \approx 0.32$ of
the system were obtained.\\


\end{abstract}

\section{Introduction}

Cataclysmic variables (CVs) are interacting binary stars which contain a
white dwarf (the primary) and an orbiting companion, usually a
main-sequence star (the secondary or the donor). The material is
accreted by the primary through the inner Lagrangian point from the
Roche-lobe filling  secondary. In non-magnetic CVs an accretion disk
is formed around the white dwarf. The region where the matter collides with
the edge of the disk is known as the hot spot (Warner 1995).

Dwarf novae (DNs) of a SU-UMa type are one of the subclasses of CVs.  
The characteristic feature of SU-UMa stars is short orbital period
(below $2.5$ h). In their light curves one can observe eruptions
classified as outbursts or superoutbursts. Outbursts to superoutbursts
occurrence ratio is ten to one. During superoutbursts SU UMa stars are
about one magnitude brighter and, in the light curves, there are
periodic light oscillations called superhumps. A superhump period is a
few percent longer than the orbital period  (more in Hellier 2001).

The period gap (from $2$ to $3$ hours) is an orbital period range where
there is a significant dearth of active CVs. The standard evolution
model assumes that around the $P_{orb} \approx 3$ hours the donor
becomes fully convective and the magnetic braking, which is responsible
for the angular momentum loss, is abruptly shut off. At that point CV
systems evolve towards shorter periods as detached binaries. When the
orbital period decreases to about 2 hours the mass transfer restarts.
The angular momentum loss is caused by the emission of gravitational
radiation (Paczy\'nski 1981) and CVs reappear as active systems below the
period gap (for a review Knigge et al. 2011). 

Up to now the variable star CzeV404 was analyzed only once;
Caga\v{s} and Caga\v{s} (2014) presented results of two observing
campaigns of CzeV404 spanning the period from June to September 2012
and June to August 2012. During that time  one superoutburst in August
2012 and a several outbursts during the 2013 observation season were
detected. Based on their observations, CzeV404 is an eclipsing
cataclismic variable of a SU UMa type.

Light curves of eclipsing dwarf novae are a reliable diagnostic tool
for the theoretical interpretation of superoutburst and superhump
mechanisms (check Smak 2013a, 2013b, B\k{a}kowska and Olech 2014),
hence our motivation for the new observing campaign of CzeV404. 

The structure of the paper is as follow: section 2 contains the
description of observation runs, data reduction and global photometric
behaviour of CzeV404. In section 3 we present the periodicity of the
detected superhumps and eclipses. Section 4 is a discussion chapter and
the last section displays summary of our results.

\bigskip

\section{Observations} 

Observations of CzeV404 reported in this work were obtained during
15 nights from 2014 June 18 to July 22 at the Borowiec station of
Pozna\'{n} Astronomical Observatory located in Poland and during 3
nights from 2014 July 27 to August 1 in Pisa in Italy. During this time
span we detected one outburst (monitored during 3 nights) and one
superoutburst (observed during 9 nights). 

Most of our observations (first 15 nights) were carried out at the
Borowiec Station. We used a 40 cm, F/4.5 Newton reflector, equipped with
a ST-7 CCD camera providing a $8.0' \times 12.0'$ field of view.

Due to unfavorable weather conditions in the last 3 nights of our
campaign, data collection was performed in Italy. For this set of
observations a Newton reflector with a $12"$ diameter, equipped with a
KAF402 CCD camera was used.

In order to obtain the shortest possible exposure times all observations
of the CzeV404 were made in the "white light" (clear filter). The
exposure times ranged from 60 to 240 seconds depending on the weather
conditions and the brightness of the object.

In total, we gathered 54.5 hours and obtained 2193 exposures of CzeV404 during 18 nights. Table 1 presents a full journal of our CCD
observations.

\subsection{Data Reduction}

We determined relative unfiltered magnitudes of CzeV404 by taking
the difference between the magnitude of the variable and the mean
magnitude of the three comparison stars. In Fig. 1 the map of a region is
displayed with the variable star marked as V1 and the comparison stars
as C1, C2 and C3, respectively. The equatorial coordinates and the
brightness of comparison star C3 (RA=$18^{h}30^{m}22^{s}.239$,
Dec=$+12^{o}32'26".17$, $12.63$ mag in $V_T$ filter) are taken from the
Tycho-2 Catalogue (Hog et al. 2000). 

\newpage

\vspace*{9.cm}

\includegraphics{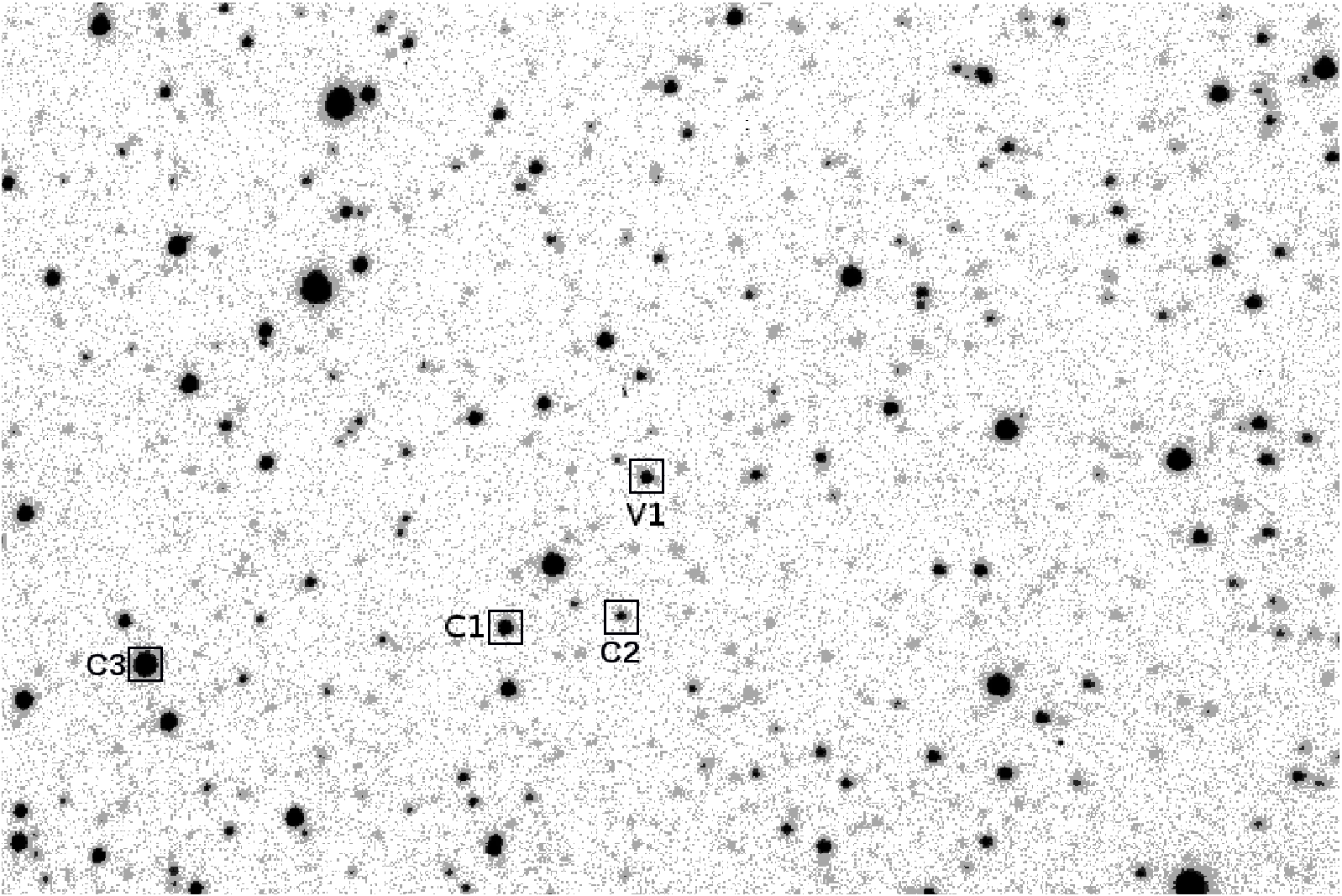}

\begin{figure}[!ht]
\caption {Finding chart of CzeV404. The variable is marked as V1.  Positions of the three comparison stars C1, C2 and C3 are also shown. The field of view is about 8.0'$\times$12.0'. North is up, east is left.}
\end{figure}

\begin{table*}
 \centering
 \begin{small}

  \caption{The journal of our CCD observations of CzeV404.}
  \vspace{0.25cm}
  \begin{tabular}{@{}|l|c|c|c|c|l|@{}}
  \hline
   Date     & Time of start   & Length   & Number of & Telescope &  Observer\\
            &  2456000+ [HJD] & of run [h] & frames  &           &          \\
\hline
    2014 June 18 & 827.37813 & 2.38 & 33  & Borowiec  & K. B\k{a}kowska \\
    2014 June 23 & 832.38466 & 3.78 & 147 & Borowiec  & K. B\k{a}kowska \\
    2014 June 26 & 835.41737 & 2.33 & 67  & Borowiec  & R. Pospieszy\'{n}ski\\
    2014 July 01 & 840.43256 & 1.25 & 73  & Borowiec  & K. B\k{a}kowska \\
    2014 July 02 & 841.36717 & 3.21 & 180 & Borowiec  & R. Pospieszy\'{n}ski\\
    2014 July 03 & 842.36747 & 1.77 & 102 & Borowiec  & R. Pospieszy\'{n}ski\\
    2014 July 04 & 843.37035 & 1.31 & 50  & Borowiec  & R. Pospieszy\'{n}ski\\
    2014 July 06 & 845.36662 & 2.31 & 90  & Borowiec  & R. Pospieszy\'{n}ski\\
    2014 July 15 & 854.42864 & 3.23 & 124 & Borowiec  & K. B\k{a}kowska \\
    2014 July 16 & 855.38603 & 1.78 & 103 & Borowiec  & K. B\k{a}kowska \\
    2014 July 17 & 856.34946 & 4.63 & 230 & Borowiec  & R. Pospieszy\'{n}ski\\
    2014 July 18 & 857.37171 & 3.43 & 199 & Borowiec  & K. B\k{a}kowska \\
    2014 July 19 & 858.34480 & 5.29 & 196 & Borowiec  & R. Pospieszy\'{n}ski\\
    2014 July 20 & 859.37679 & 4.32 & 154 & Borowiec  & A. Marciniak  \\
    2014 July 22 & 861.36142 & 4.78 & 180 & Borowiec  & A. Marciniak   \\
    2014 July 27 & 866.40934 & 4.60 & 140 & Pisa      & F. Martinelli \\
    2014 July 31 & 870.36370 & 1.83 &  51 & Pisa      & F. Martinelli \\
    2014 August 1 & 871.34353 & 2.31 & 74  & Pisa      & F. Martinelli \\

\hline
\end{tabular}
 \end{small}
\end{table*}

\newpage
We performed all of the data reductions using the IRAF\footnote{IRAF is
distributed by the National Optical Astronomy Observatory, which is
operated by the Association of Universities for Research in Astronomy,
Inc., under a cooperative agreement with the National Science
Foundation.} package. Profile photometry was obtained with DAOPHOTII (Stetson 1987). The accuracy of our measurements varied between 0.006 and 0.085 mag for
observations from Poland, 0.006 and 0.045 mag for the data set from
Italy,  depending on the atmospheric conditions and brightness of the
variable. The median value of the photometry errors was 0.010  mag and
0.013 mag for Borowiec and Pisa observations, respectively.

\subsection{Light curves}

From now on we use only a day number HJD-2456000 [d] to
refer to our observations. In Fig. 2 one can see the global light curve
of CzeV404 from 2014 June 18 to August 1. 
 
In the begining of our observations (HJD $827-835$) CzeV404 was in
quiescence with brightness varying between $V_T \approx 16.6$ and $V_T
\approx 17.0$ mag. Later (HJD $840-842$), we caught the variable during
the outburst with the maximum brightness of $V_T \approx 15.5$ mag. Due
to lack of observations between HJD $836-839$, caused by weather
conditions, we can only say that the outburst lasted at least 3 nights
and the amplitude was not lower than $A_o \approx 1.5$ mag. Based on
these estimate, it is highly probable that we observed almost the whole outburst.
On HJD $854$ the superoutburst started and lasted for 17 nights. During
the superoutburst the brightness of CzeV404 reached $V_T \approx
15.0 $ mag at maximum and decreased to $V_T \approx 17.2 $ mag at
minimum, resulting in the amplitude of $A_s \approx 2.2 $ mag. 

In Fig. 3 we present the individual light curves from four nights HJD
$856-861$ of the variable during its July 2014 superoutburst. Short-term
modulations known as superhumps are visible during all of the four
displayed nights.

\vspace*{10.0cm}
\includegraphics{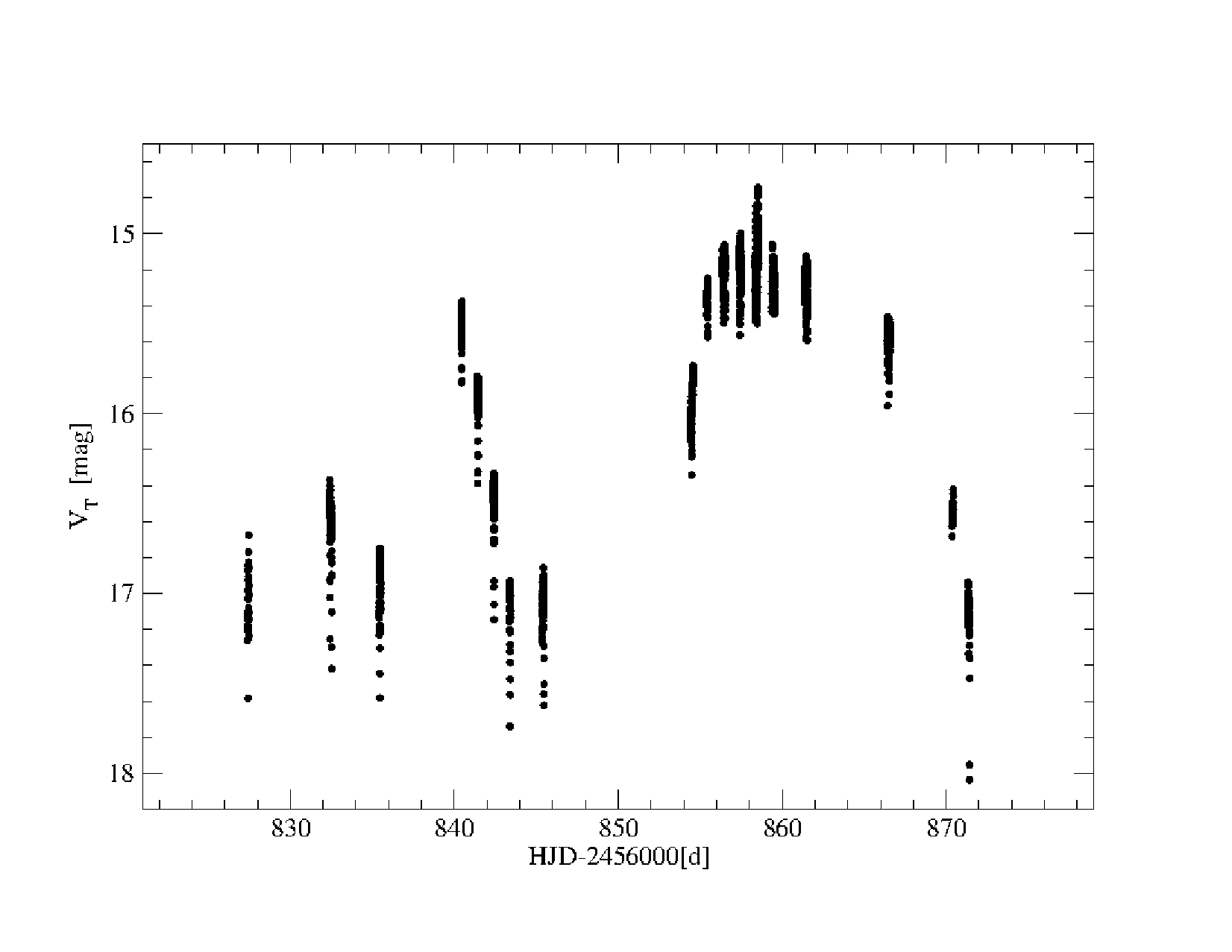}

   \begin{figure}[!ht]
      \caption {Global photometric behaviour of CzeV404 during our observational campaign in June-August 2014.}
   \end{figure}  

\vspace*{10.0cm}

\includegraphics{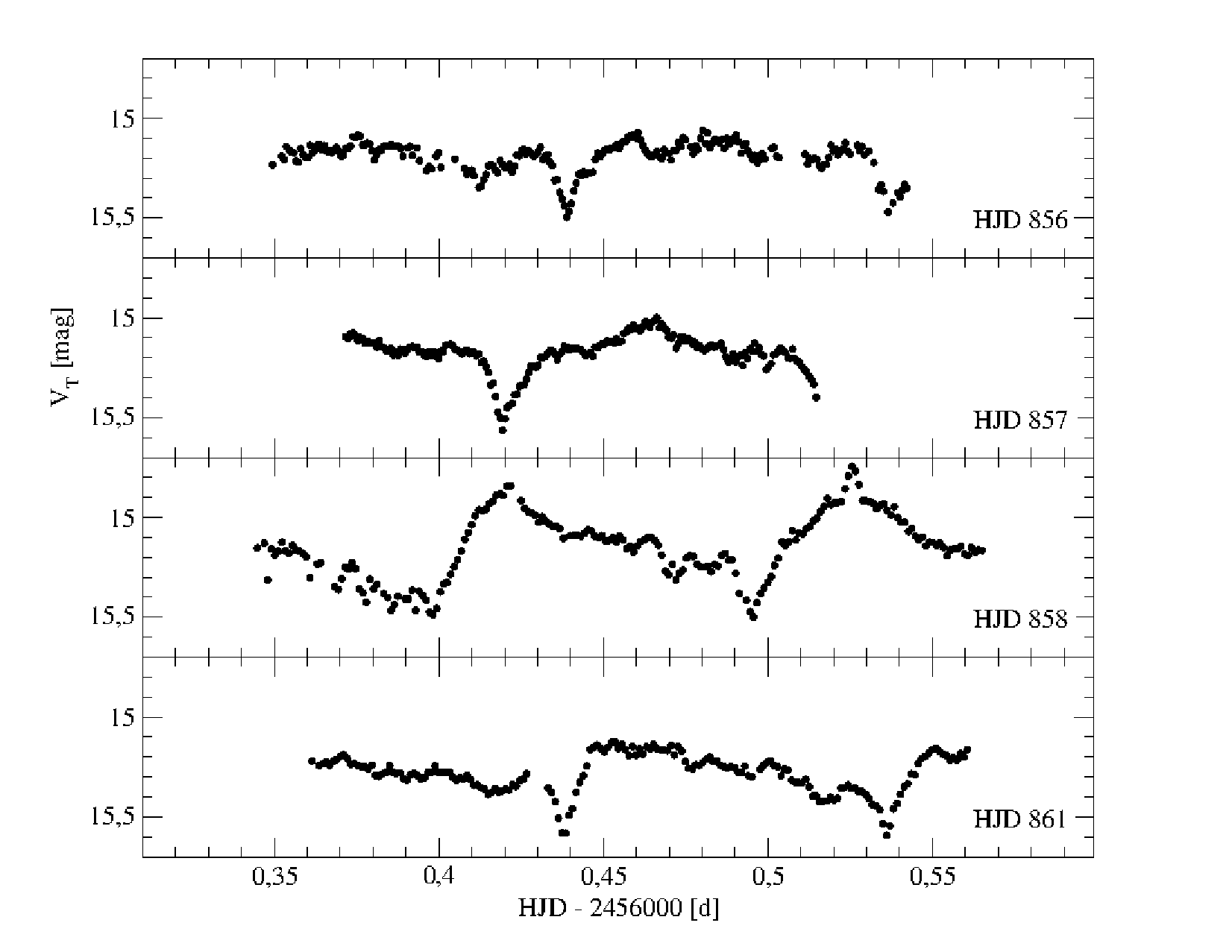}

\begin{figure}[!ht]
\caption {Light curves of CzeV404 during its superoutburst in July 2014. A fraction of HJD is presented on the $x$-axis. Additionally, HJD$-2456000$ [d] is given on the right  side of each panel.}
\end{figure} 
 
\section{Superhumps and Orbital Periods}

\subsection{$O-C$ Diagram for Superhumps}

To calculate the period of light modulations during the superoutburst we
constructed the $O-C$ diagram for the times of superhumps maxima. In
total, we determined nine peaks of maxima, which are listed in Table 2
together with their errors, cycle numbers $E$ and $O-C$ values.

The least squares linear fit to the data presented in Table 2 allowed to
obtain the following ephemeris:

\begin{equation}
{\rm HJD_{\rm max}} = 2456856.4286(5) + 0.10472(2) \times E
\end{equation} 

\noindent which indicates that the value of the superhump period is
$P_{sh}=0.10472(2)$ days ($ 150.80 \pm 0.03$ min).

In Fig. 4 we display the $O-C$ values
corresponding to the ephemeris (1). 

The decreasing trend of the superhump period shown in Fig. 4 was
confirmed by calculations of the second-order polynomial fit to the
moment of maxima. The following ephemeris was obtained:

\begin{equation}
{\rm HJD_{\rm max}} = 2456856.4055(9) + 0.10618(5) \times E - 1.29(4) \times 10^{-5} \times E^{2} 
\end{equation}

In Fig. 4 the solid line corresponds to the ephemeris (2). 

After this investigation, we conclude that the period of superhumps was
not stable during the July 2014 superoutburst of CzeV404 and it can
be described by a decreasing trend with a rate of
\textit{\.{P}}=$-2.43(8) \times 10^{-4}$. 

\begin{small}
\begin{table*}[!h]
\caption{Times of superhumps maxima in the light curves of CzeV404 during  its July 2014 superoutburst.}
\vspace{0.1cm}
\begin{center}
\begin{tabular}{|r|c|r|r|}
\hline
\hline
$E$ & ${\rm HJD}_{\rm max} - 2456000$ & Error & $O-C$ \\
    &                                 &       & [cycles]\\
\hline
0   & 856.3774 & 0.0020 & -0.4885  \\ 
1   & 856.4911 & 0.0020 & -0.4026  \\
10  & 857.4666 & 0.0015 & -0.0855  \\
19  & 858.4215 & 0.0005 &  0.0349  \\
20  & 858.5260 & 0.0005 &  0.0330  \\
48  & 861.4652 & 0.0010 &  0.1055  \\
95  & 866.3655 & 0.0020 & -0.0912  \\
97  & 866.5715 & 0.0030 & -0.1237  \\
134 & 870.4270 & 0.0030 & -0.2994  \\
\hline
\end{tabular}
\end{center}
\end{table*}
\end{small}

\vspace*{10.0cm}
\includegraphics{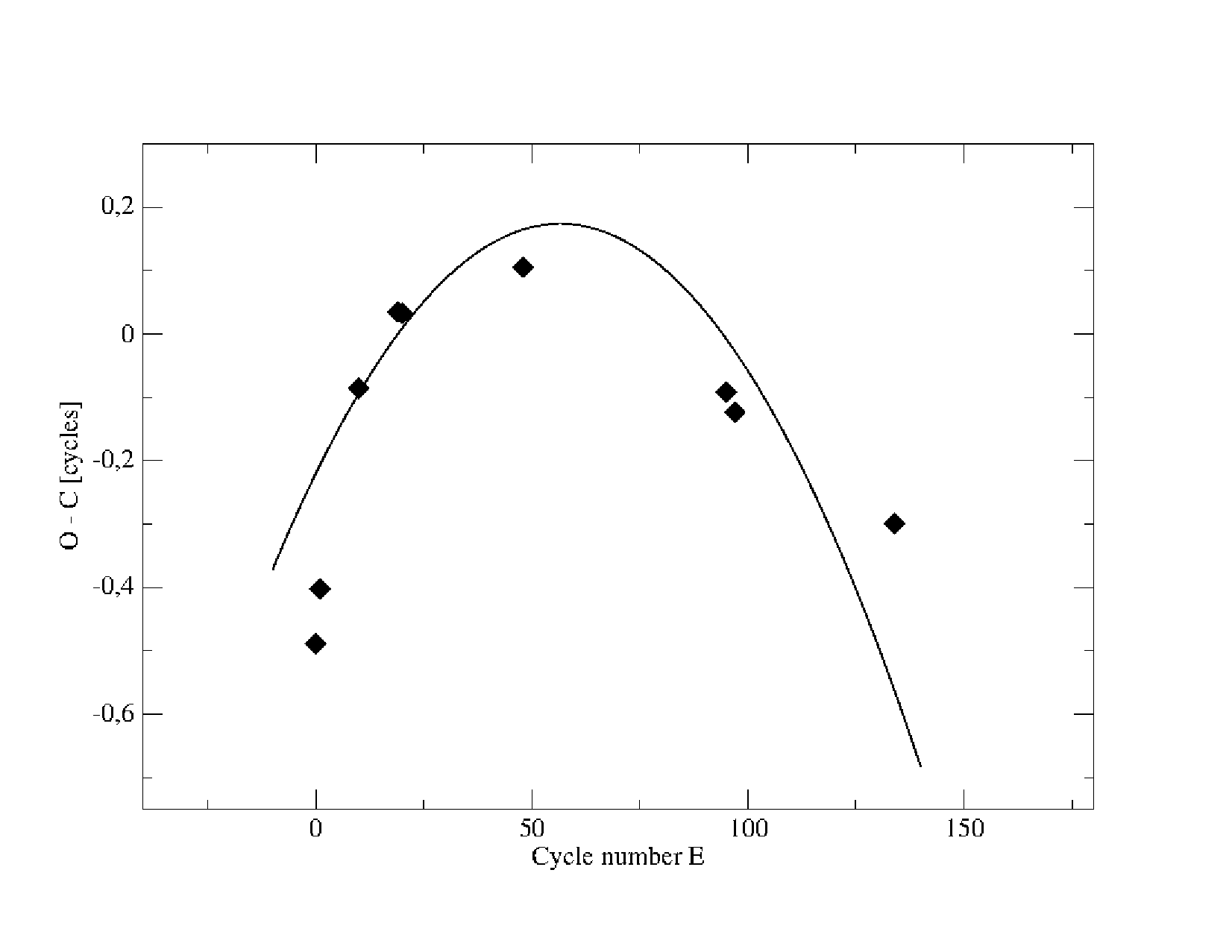}

\begin{figure}[!ht]
\caption {The $O-C$ diagram for the superhumps maxima of CzeV404 observed during its superoutburst in July 20. The solid line corresponds to the fit presented in Eq.(2).}
\end{figure} 

\subsection{$O-C$ Diagram for Eclipses}

We obtained the value of the orbital period by constructing the $O-C$
diagram for the moments of the minima. In total, we used the timings of
17 eclipses from the June-August 2014 observing season and the following
ephemeris of the minima was derived: 

\begin{equation}
{\rm HJD_{\rm min}} = 2456827.4249(1) + 0.0980203(6) \times E
\end{equation} 

\noindent which gives the orbital period of $P_{orb1}=0.0980203(6)$
days (141.1492 $ \pm 0.0009$ min).

In Table 3 we present the timings of eclipses with errors, cycle numbers
E and $O-C$ values. The resulting $O-C$ diagram for the moments of
minima is shown in Fig. 5. 

Between the superhump and orbital periods there is a relation introduced by Osaki (1985):

\begin{equation}
{\frac{1}{P_{sh}} = \frac{1}{P_{orb}} - \frac{1}{P_{beat}}},
\end{equation}

\noindent and we used Eq.(4) and the orbital $P_{orb1}$ and superhump
$P_{sh}$ periods to calculate the beat period of $P_{beat}=1.53 \pm
0.02$ days. 

To obtain the best possible value of the orbital period we combined our
17 timings of eclipses from 2014 and 16 from the 2012-2013 observations
presented by Caga\v{s} and Caga\v{s} (2014). Based on this, we
calculated the following ephemeris of the minima:

\begin{equation}
{\rm HJD_{\rm min}} = 2456827.42359(5) + 0.09802201(2) \times E
\end{equation} 

\noindent and this corresponds to the orbital period of
$P_{orb2}=0.09802201(2)$ days ($ 141.15169 \pm 0.00003 $ min). In Fig. 6
we show the resulting $O-C$ diagram for the moments of eclipses for
2012-2014 time span.

\begin{small}
\begin{table*}[!h]
\caption{Times of minima in the light curves of CzeV404 observed in the period of June-July 2014.}
\vspace{0.1cm}
\begin{center}
\begin{tabular}{|r|c|r|r|}
\hline
\hline
$E$ & ${\rm HJD}_{\rm min} - 2456000$ & Error & $O-C$ \\
    &                                 &       & [cycles]\\
\hline
0    & 827.4253 & 0.0005 &  0.0037 \\ 
51   & 832.4242 & 0.0002 &  0.0023 \\
52   & 832.5219 & 0.0002 & -0.0010 \\
82   & 835.4628 & 0.0004 &  0.0020 \\
133  & 840.4615 & 0.0005 & -0.0014 \\
143  & 841.4417 & 0.0002 & -0.0014 \\
153  & 842.4216 & 0.0003 & -0.0045 \\
163  & 843.4024 & 0.0005 &  0.0016 \\
184  & 845.4604 & 0.0005 & -0.0028 \\
286  & 855.4583 & 0.0004 & -0.0045 \\ 
296  & 856.4388 & 0.0003 & -0.0015 \\
297  & 856.5370 & 0.0003 &  0.0004 \\
306  & 857.4194 & 0.0002 &  0.0026 \\
317  & 858.4969 & 0.0005 & -0.0048 \\
347  & 861.4380 & 0.0003 &  0.0002 \\
348  & 861.5361 & 0.0003 &  0.0010 \\
449  & 871.4362 & 0.0005 &  0.0015 \\
\hline
\end{tabular}
\end{center}
\end{table*}
\end{small}

\section{Discussion}

Knowing the orbital and superhump periods we can draw conclusions about
the evolution of CzeV404. In the diagram $P_{orb}$ versus
$\varepsilon$ presented in Fig. 7 we display SU UMa stars, period
bouncers and nova-like variables. The period excess was calculated as:

\begin{equation}
\varepsilon = \frac{\Delta P}{P_{orb}} =\frac{P_{sh}-P_{orb}}{P_{orb}}.
\end{equation}

\noindent and for CzeV404 we obtained  a value of $\varepsilon=
6.8\% \pm 0.02 \%$.

\newpage

\vspace*{8.0cm}
\includegraphics{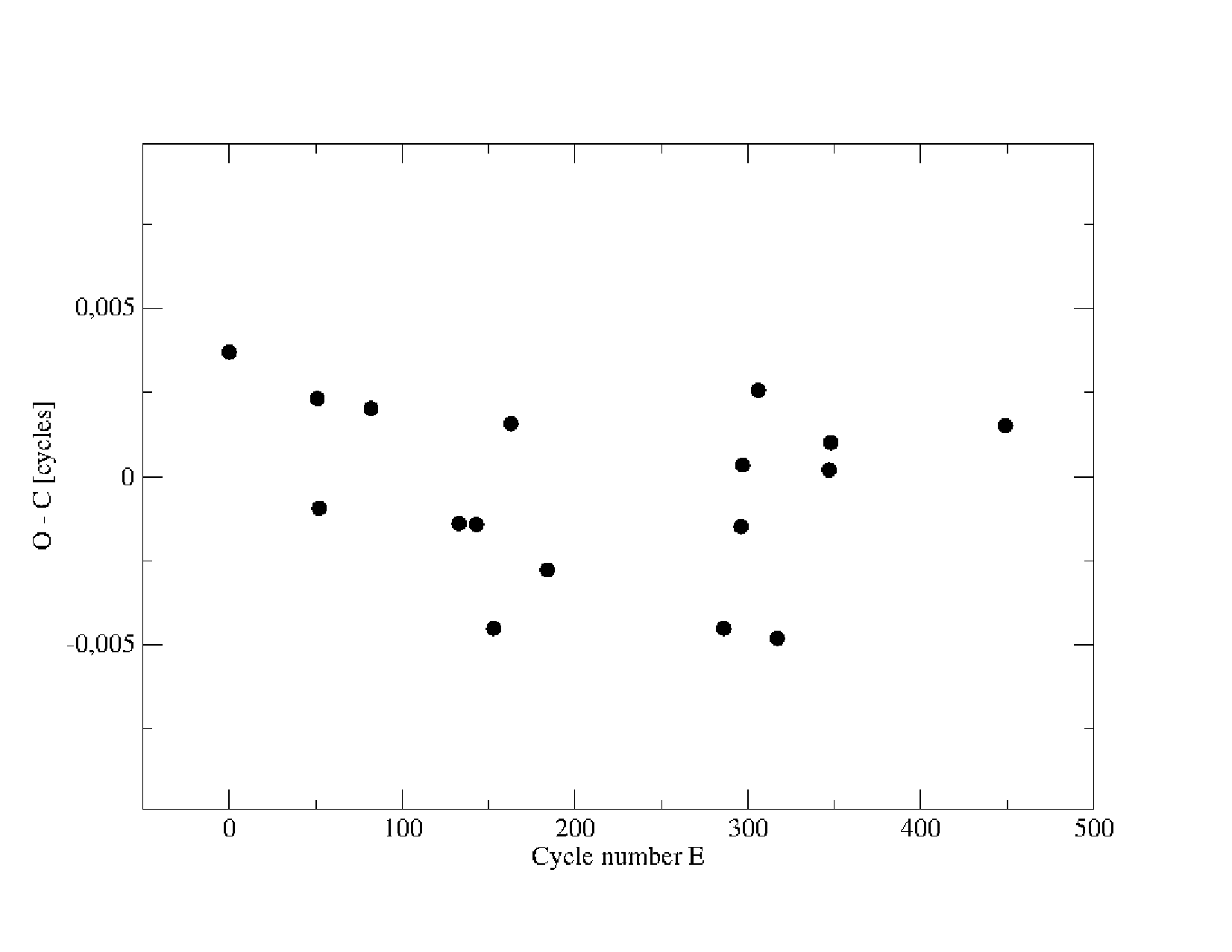}

\begin{figure}[!ht]
\caption {The $O-C$ diagram for the moments of eclipses observed in CzeV404 during the June-August 2014 campaign.}
\end{figure} 

\vspace*{8.0cm}
\includegraphics{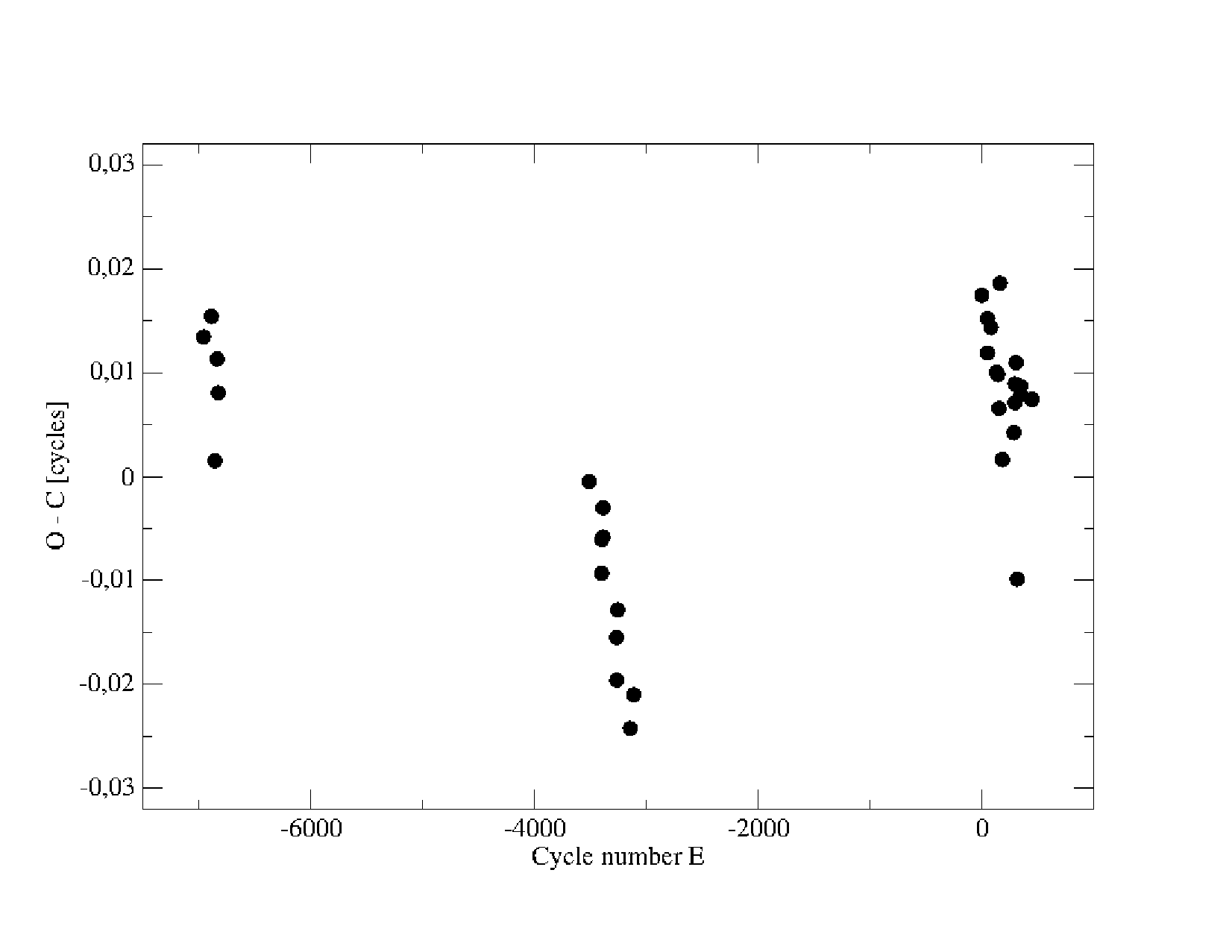}

\begin{figure}[!ht]
\caption {The $O-C$ diagram for the moments of minima was based on our data set from 2014 and the data from 2012-2013 provided by Caga\v{s} and Caga\v{s} (2014).}
\end{figure} 

\vspace*{10.0cm}
\includegraphics{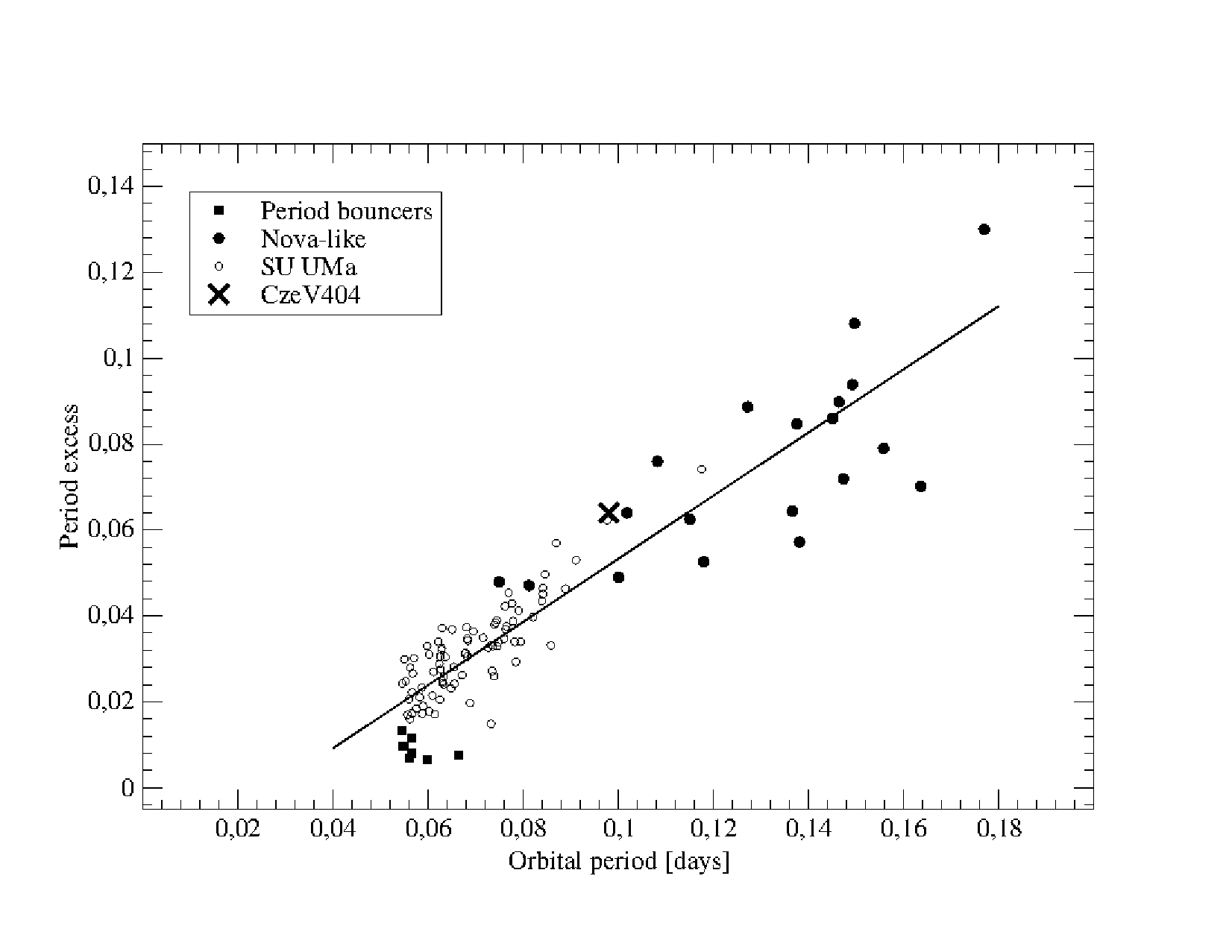}

\begin{figure}[!ht]
\caption {The relation between the period excess and the orbital period for several subclasses of CVs. CzeV404 is marked by a big, black $X$-mark. By open circles we plotted SU UMa type stars. Nova-like variables are illustrated by black dots. The period bouncer candidates are represented by black squares (figure taken from Olech et al. 2011).}
\end{figure} 

\newpage

The mass ratio of the binary $q=M_2/M_1$ can be obtained from the 
empirical formula introduced by Patterson (1998): 

\begin{equation}
\varepsilon = \frac{0.23q}{1+0.27q}.
\end{equation}

\noindent This results with the $q \approx 0.32$ for CzeV404. The
tidal instability (Whitehurst  1988) of the disk starts to work
effectively for binaries with a mass ratio $q$ below $0.25$. This
assumption was used by Osaki (1989) in the TTI (thermal-tidal
instability) model to explain the phenomenon of superoutbursts and
superhumps. That is why such a high value of $q$ in CzeV404 poses a serious problem for the superhump mechanism. 

The detection of active DNs in the period gap, for example TU Men (Stolz and Schoems 1981),  MN Dra (Nogami et
al. 2003), NY Ser (Pavlenko et al. 2010, 2014), SDSS J162520.29+120308.7
(Olech et al. 2011), OGLE-BLG-DN-001 (Poleski et al. 2011) is another
issue in the evolution of CVs. There are two mechanisms which are
thought to drive the orbital angular momentum loss. First one is the
magnetic wind breaking which is greatly reduced around orbital period
$P_{orb} \approx 3$ hours when the secondary becomes fully convective
(Verbunt and Zwaan 1981) and the donor detaches from its Roche-lobe.
Between the $2-3$ hour orbital period range, know as the orbital period
gap, the number of CVs found is very low (see G\"{a}nsicke et al. 2009).
At around $P_{orb} \approx 2$ hours (Howell et al. 2001) the mass
transfer is resumed because the secondary reestablished the Roche lobe
contact. Further evolution of CVs is driven mainly by the second
mechanism of the orbital angular momentum loss known as the
gravitational radiation (Paczy\'{n}ski 1981). SU UMa stars located in
period gap and having Roche lobe filling secondaries should be driven by
gravitational wave radiation. At this orbital period ($\sim 2.5$ hours)
both the angular momentum loss as well as the mass transfer should be low and according
to the TTI model these objects should be characterized by low activity.
It is certainly not the case in such systems as CzeV404 or MN Dra. 

The first suggestion that superhump period in majority of SU UMa stars
changes with the common pattern appeared in Olech et al. (2003, 2004).
Since 2009, Kato et al. (2009, 2010, 2012. 2013, 2014a, 2014b) published 
extensive surveys concerning superhump period changes observed in SU
UMa stars. According to these works the evolution of superhump period can
be divided into three parts: stage A with a stable and longer period,
stage B characterized by a positive period derivative and the last stage C
with a shorter and stable period.  Several active DNs in the period gap,
i.e. SDSS J170213 (Kato et al. 2013), CSS J203937, V444 Peg, MASTER
J212624 (Kato et al. 2014a) and MN Dra (Kato et al. 2014b) were presented
in surveys supervised by the Japanese team and a part of them followed this
scenario. However, some of the long period SU UMa stars seem to show only
a decreasing superhump period across the entire superoutburst (see for  example
SDSS J1556 and UV Per in Kato et al. 2009). CzeV404 seems to belong
to the same group. Our observations cover the entire superoutburst and there
is a clear and a rather constant superhump period decrease observed over
the whole interval of 17 days.

It is worth noting that some of the CzeV404 system parameters are
similar to MN Dra, respectively, i.e. orbital periods
$P_{orb}=0.0998(2)$ days (Pavlenko et al. 2010) and
$P_{orb}=0.0980203(6)$ days (this work), superhump periods
$P_{sh}=0.105040(66)$ days (Kato et al. 2014b) and $P_{sh}=0.10472(2)$ days (this
work), the large period variations \textit{\.{P}}=$-1.48(95) \times
10^{-4}$ (Kato et al. 2014b) and \textit{\.{P}}=$-2.43(8) \times 10^{-4}$
(this work), and the mass ratio $q \approx 0.29$ (Kato et al. 2014b) and
$q \approx 0.32$ (this work). That is why both of these active DNs from
the period gap deserve more observations for further analysis of
CV stars evolution. 

\section{Conclusions}

We summarize the results of the summer 2014 campaign of CzeV404:

\begin{itemize}

\item During the two and half months of observations from 2014 June 18 to
August 1 we detected one outburst and one superoutburst in CzeV404.
The outburst lasted at least 3 nights and the amplitude was not fewer
then $A_o \approx 1.5 $ mag.  The June 2014 superoutburst had the
amplitude of brightness $A_s \approx 2.2 $ mag and had a duration of 17
nights. Clear superhumps were detected during the superoutburst.

\item Based on the maxima of superhumps we determined the superhump
period $P_{sh}=0.10472(2)$ days ($ 150.80 \pm 0.03$ min). The orbital
period $P_{orb1}=0.0980203(6)$ days (141.1492 $ \pm 0.0009$ min) was obtained 
using 17 eclipses. Combining our data set and the moments of
eclipses provided by Caga\v{s} and Caga\v{s} (2014) we found a more
precise orbital period with the value of $P_{orb2}=0.09802201(2)$ days
($ 141.15169 \pm 0.00003 $ min).

\item The superhump period was not stable during the July 2014
superoutburst of CzeV404 and it was decreasing with the high value 
of \textit{\.{P}}=$-2.43(8) \times 10^{-4}$. 

\item With the Stolz-Schoembs (1984) formula we derived period excess
$\varepsilon= 6.8\% \pm 0.02 \%$ and on the diagram $P_{orb}$ versus
$\varepsilon$ CzeV404 is located in the period gap objects. 

\item The mass ratio with the value of $q \approx 0.32 $ was obtained.
Such a high value is a problem for the TTI model assuming $q < 0.25 $
for the superhump and the superoutburst mechanism. 

\end{itemize}

There is no doubt that CzeV404 is a unique variable star. It belongs
to about one dozen of SU UMa stars located in the period gap. Its
orbital period is one of the longest among whole group of SU UMa
variables. Additionally, it is longest period eclipsing SU UMa star
known and its high mass ratio poses a serious challenge for the TTI model.
Certainly, CzeV404 should be studied further. Both high speed photometry
with 2-meter class telescope as well as spectroscopy with the world's largest
telescopes would allow to precisely determine the basic
parameters of the binary system.

\bigskip 

\noindent {\bf Acknowledgments.} ~We acknowledge generous allocation of Pozna\'{n} Observatory 0.4-m 
telescope time. Also, we want to thank mgr A. Borowska for language corrections. Project was supported by  Polish
National Science Center grant awarded by decision
DEC-2012/07/N/ST9/04172 for KB.


\begin{thebibliography}{}

\bibitem{Bakowska14c} B\k{a}kowska K. and Olech A.,  2014, {\it Acta Astron.}, 64, 247

\bibitem{Cagas} Caga\v{s} P. and Caga\v{s} P., 2014, {\it IBVS}, 6097, 1

\bibitem{Gansicke09} G\"{a}nsicke B.T., Dillon M., Southworth J. et al., 2009, {\it MNRAS}, 397, 2170

\bibitem{Hell00} Hellier C., 2001,  {\it Cataclysmic Variable Stars}, Springer

\bibitem{Hog00} Hog E., Fabricius C., Makarov V.V. et al., 2000, {\it A\&A}, 355, L27 

\bibitem{Howell01} Howell S.B., Nelson L.A. and Rappaport S., 2001, {\it ApJ}, 550, 897

\bibitem{Knigge11} Knigge C., Baraffe I. and Patterson J., 2011, {\it ApJS}, 194, 28

\bibitem{Kato09} Kato T., Imada A., Uemura M. et al., 2009, {\it PASJ}, 61, S395

\bibitem{Kato10} Kato T., Maehara H., Uemura M. et al., 2010, {\it PASJ}, 62, 1525

\bibitem{Kato12} Kato T., Maehara H., Miller I. et al., 2012, {\it PASJ}, 64, 21

\bibitem{Kato13a} Kato T., Hambsch F.-J., Maehara H. et al., 2013a, {\it PASJ}, 65, 23

\bibitem{Kato13b} Kato T., Hambsch F.-J., Maehara H.  et al., 2014a, {\it PASJ}, 66, 30

\bibitem{Kato14} Kato T., Dubovsky P.A., Kudzej I. et al., 2014b, {\it arXiv:1406.6428},

\bibitem{Nogami03} Nogami D., Uemura M., Ishioka R. et al., 2003, {\it A\&A}, 404, 1067

\bibitem{Olech03} Olech A., Schwarzenberg-Czerny A., K\k{e}dzierski P., Z{\l}oczewski K., Mularczyk K., Wi\'{s}niewski M., 2003, {\it Acta Astron.}, 53, 175

\bibitem{Olech04} Olech A., Cook L.M., Z{\l}oczewski K., Mularczyk K., K\k{e}dzierski P., Udalski A., Wi\'{s}niewski M., 2004, {\it Acta Astron.}, 54, 233

\bibitem{Olech00} Olech A., de Miguel E., Otulakowska M. et al. 2011, {\it A\&A}, 532, A64

\bibitem{Osaki85} Osaki Y., 1985, {\it A\&A}, 144, 369

\bibitem{Osaki89} Osaki Y., 1989, {\it PASJ}, 41, 1005

\bibitem{Paczynski81} Paczy\'{n}ski B., 1981, {\it Acta Astron.}, 31, 1

\bibitem{Patt98} Patterson J., 1998, {\it PASP}, 110, 1132

\bibitem{Pavlenko10} Pavlenko E., Kato T., Andreev M. et al., 2010, {\it 17th European White Dwarf Workshop AIP Conference Proceedings}, 1273, 320 

\bibitem{Pavlenko14} Pavlenko E.P., Kato T., Amtonyuk O.I. et al., 2014, arXiv:1408.4285) 

\bibitem{Poleski11} Poleski R., Udalski A., Skowron J. et al., 2011, {\it Acta Astron.}, 61, 123


\bibitem{Stolz84} Stolz B. and Schoembs R., 1981, {\it IBVS}, 1955, 0374 

\bibitem{Stolz84} Stolz B. and Schoembs R., 1984, {\it A\&A}, 132, 187 

\bibitem{Smak13a} Smak J., 2013a, {\it Acta Astron.}, 63, 109

\bibitem{Smak13b} Smak J., 2013b, {\it Acta Astron.}, 63, 369

\bibitem{Stetson87} Stetson P.B., 1987, {\it PASP}, 99, 191

\bibitem{Verbunt1981} Verbunt F. and Zwaan C., 1981, {\it A\&A}, 100, L7

\bibitem{war95} Warner B., 1995, {\it Cataclysmic Variable Stars}, 
           Cambridge University Press

\bibitem{Whitehurst88} Whitehurst R., 1988, {\it MNRAS}, 232, 35

\end{thebibliography}
\end{document}